\documentclass[a4paper, 11pt]{article}
\usepackage{srcltx}
\usepackage{indentfirst}
\usepackage{graphicx}
\usepackage{overpic}
\usepackage{multirow}
\usepackage[dvipdfmx]{hyperref}
\usepackage[hang]{subfigure}
\usepackage{amsmath, amssymb, amsthm}
\usepackage{color}
\usepackage[mathscr]{euscript}

\newcommand\bbR{\mathbb{R}}
\newcommand\bbN{\mathbb{N}}
\newcommand\bxi{\boldsymbol{\xi}}
\newcommand\bx{\boldsymbol{x}}
\newcommand\bv{\boldsymbol{v}}
\newcommand\bu{\boldsymbol{u}}

\newcommand\dd{\,\mathrm{d}}
\newcommand\He{\mathit{He}}
\newcommand\Kn{\mathit{Kn}}

\newcommand\Rxx{R$xx$~}
\newcommand\NRxx{NR$xx$~}

\numberwithin{equation}{section}

\graphicspath{{images/}}

{\theoremstyle{remark} }

\title{An Efficient \NRxx Method for Boltzmann-BGK Equation}

\author{Zhenning Cai\thanks{School of Mathematical Sciences, Peking
    University, Beijing, China, email: {\tt harecat@gmail.com}.},~~ Ruo
  Li\thanks{CAPT, LMAM \& School of Mathematical Sciences, Peking
    University, Beijing, China, email: {\tt rli@math.pku.edu.cn}.}~~ and
  Yanli Wang\thanks{School of Mathematical Sciences, Peking
    University, Beijing, China, email: {\tt wangyanliwyl@gmail.com}.}}

\begin{document}
\maketitle
\begin{abstract}
  In \cite{NRxx}, we proposed a numerical regularized moment method of
  arbitrary order (abbreviated as \NRxx method) for Boltzmann-BGK
  equation, which makes numerical simulation using very large number
  of moments possible. In this paper, we are further exploring the
  efficiency of \NRxx method with techniques including the 2nd order
  HLL flux with linear reconstruction to improve spatial accuracy, the
  RKC schemes to relieve the time step length constraint by the
  regularization terms, and the revised Strang splitting to calculate
  convective and diffusive terms only once without loss of
  accuracy. It is validated by the numerical results that the overall
  efficiency is significantly improved and the convergence order is
  kept well.
\end{abstract}

\section{Introduction}
In 1949, the moment method was introduced by Grad \cite{Grad} as a
technique to approximate the Boltzmann equation in a macroscopic view.
With proficient mathematical skills, he derived the famous 13-moment
system, but the system is problematic: the hyperbolicity can only be
obtained in the neighbourhood of Maxwellian, and the structure of
shock wave is non-smooth when the Mach number is large (see e.g.
\cite{Muller, Grad1952}). During a long time, the moment method
suffers lots of criticism, and very few progresses are made before
1990s. In the recent twenty years, a number of new ideas based on the
moment method have come out, such as the Jin-Slemrod regularization of
the Burnett equations \cite{Jin}, the COET (Consistently Ordered
Extended Thermodynamics) method \cite{Reitebuch}, the order of
magnitude approach \cite{Struchtrup2005}, and et al. In
\cite{Struchtrup2003}, regularization based on 1st order
Chapman-Enskog expansion was considered to remedy the defects of
Grad's 13 moment equations. The regularized system obtained therein
was referred as R13 equations. Soon, the R20 and R26 equations are
studied \cite{Mizzi,Emerson} and the aspiration to extend this method
for system with more moments \cite{Struchtrup2008} was called as \Rxx
by the authors of \cite{Struchtrup2003}. In \cite{NRxx}, we proposed a
numerical regularized moment method of arbitrary order for
Boltzmann-BGK equation. The method is abbreviated as ``\NRxx method''
later on for convenience following the tradition.  The \NRxx method
makes it possible to investigate large moment systems, by direct
numerical approximation of the regularized moment systems without
deriving the explicit forms of the moment equations. The
regularization in \cite{NRxx} is similar to the original derivation of
the R13 equations \cite{Struchtrup2003}. And later in \cite{NRxx_new},
the idea of COET \cite{Reitebuch} is adopted to revise the
regularization terms, which results in a parabolic system.

Though one can escape from deriving the complex moment system, and the
developing of the simulation program is greatly simplified by the
\NRxx method, the numerical efficiency of the method in \cite{NRxx}
should be further explored. It has been verified that the
computational time is linear in the number of moments. However, a 1st
order HLL numerical flux, which is over diffusive, was used in
\cite{NRxx} for the transportation part, such that we have to use quite
fine spatial grids to reduce the numerical error to a moderate level.
Noting that the regularized moment system is parabolic, the time step
is quadratic in the grid size. It turns out that in the 1D case, the
total computational cost is cubic in the number of spatial grids. With
a quite fine spatial grid to achieve enough accuracy, the numerical
simulation therein is rather time consuming.

In this paper, we focus on improving the efficiency of the \NRxx
method in \cite{NRxx} with some state-of-art techniques available to
us. Precisely, three numerical techniques are employed:
\begin{enumerate}
\item \emph{Linear spatial reconstruction. } The piecewise linear
  spatial reconstruction is able to provide a 2nd order HLL flux and
  greatly reduces the numerical diffusion. Owing to the absence of
  analytical expressions of the moment equations, the reconstruction
  needs to be done carefully. A conservative reconstruction is
  proposed, and it is numerically verified to have achieved high
  resolution. Although the final scheme is still of first order, the
  numerical error is greatly reduced. Thus much less grids are used
  in the computation.  
\item \emph{RKC time integration scheme.} The RKC time integration
  scheme \cite{RKC} is adopted to enlarge the time step sizes. The RKC
  method is a Runge-Kutta type method originally designed for 
  diffusive PDEs to provide large stability regions, while the
  regularized moment equations are convective-diffusive problems,
  where imaginary parts appear in the eigenvalues of the right
  hand side of the semi-discrete system. As is well known, the RKC
  schemes contain a parameter called as the damping factor, which
  allows a small imaginary perturbation of eigenvalues and is usually
  selected as a small positive value. In order to make the RKC scheme
  compatible with the convection terms, we use a large damping factor
  in our scheme to ensure stability. The final time step size is
  equivalent to the grid size, and the number of internal time steps
  is equivalent to the square root of the grid number. Thus the total
  number of time steps is essentially reduced, and no instability
  phenomenon apprears in our numerical test.
\item \emph{Revised Strang splitting method.} With enlarged time step,
  the numerical accuracy can be harmed by convection-collision
  splitting. Therefore, the Strang splitting is utilized to win higher
  order of accuracy in the time direction. Usually, the implementation
  of the Strang splitting requires twice calculations of the collision
  term in one time step.  For this problem, we combine the two steps
  of collision in the successive time steps, so only once calculation
  of the collision term is performed in a time step.
\end{enumerate}
Several numerical examples are carried out to show the efficiency of
our algorithm. Our prediction of the computational cost together with
the order of convergence is validated by numerical experiments. We
also compare the current method with the scheme without linear
reconstruction to demonstrate much higher resolution for the one with
linear reconstruction.

The rest of this paper is arranged as follows: in Section
\ref{sec:model}, we give a brief review of the Boltzmann-BGK equation
and the \NRxx method. In Section \ref{sec:reconstruction}, the linear
reconstruction is added to the HLL numerical flux. In Section \ref
{sec:time_step}, the time step size is enlarged by the RKC method, and
the accuracy is improved by the Strang splitting method. In Section
\ref{sec:num_ex}, numerical examples are carried out to make
illustrations. Some concluding remarks are given in Section \ref
{sec:conclusion}.


\section{The \NRxx method} \label{sec:model}
\subsection{The Boltzmann-BGK model}
In the kinetic theory, it is generally accepted that the Boltzmann
equation is able to describe the fluids accurately. Due to the
complexity of the collision term, in the computational field, several
simplified collision operators are adopted, among which the BGK
model \cite{BGK} is the simplest but useful. The Boltzmann-BGK
equation reads
\begin{equation} \label{eq:BGK}
\frac{\partial f}{\partial t} +
  \bxi \cdot \nabla_{\bx} f = \nu (f - f_M),
\qquad \bx, \bxi \in \bbR^D,
\end{equation}
where $f$ is the distribution function, $\nu$ is the collision
frequency, and $f_M$ is the local Maxwellian defined by
\begin{equation}
f_M(t, \bx, \bxi) = \frac{\rho(t, \bx)}{[2\pi \theta(t, \bx)]^{D/2}}
  \exp \left(-\frac{|\bxi - \bu(t, \bx)|^2}{2\theta(t, \bx)} \right).
\end{equation}
Here $\rho$, $\bu$ and $\theta$ are local macroscopic variables which
represent the density, velocity and temperature respectively. They can
be calculated from the distribution function $f$ by
\begin{align}
\rho(t, \bx) &= \int_{\bbR^D} f(t,\bx,\bxi) \dd \bxi, \\
\rho(t, \bx) \bu(t, \bx) &= \int_{\bbR^D} \bxi f(t,\bx,\bxi) \dd \bxi, \\
\rho(t, \bx) \bu(t, \bx) + D \rho(t, \bx) \theta(t, \bx)
  &= \int_{\bbR^D} |\bxi|^2 f(t,\bx,\bxi) \dd \bxi.
\end{align}

\subsection{A first-order scheme for the \NRxx method} \label{sec:NRxx}
The \NRxx method is raised in \cite{NRxx, NRxx_new} as a new tool for
the computation of large moment systems. It is based on an Hermite
expansion of the distribution function:
\begin{equation} \label{eq:expansion}
f(t, \bx, \bxi) = \sum_{\alpha \in \bbN^D} f_{\alpha}(t, \bx)
  \mathcal{H}_{\theta(t,\bx), \alpha} \left(
    \frac{\bxi - \bu(t, \bx)}{\sqrt{\theta(t, \bx)}}
  \right),
\end{equation}
where $\alpha$ is a $D$-dimensional multi-index. $\mathcal{H}_{\theta,
\alpha}$'s and $f_{\alpha}$'s act as the basis functions and the
corresponding coefficients respectively, and $\mathcal{H}_{\theta,
\alpha}$ is defined by
\begin{equation}
\mathcal{H}_{\theta,\alpha}(\bv) =
  (2\pi)^{-\frac{D}{2}} \theta^{-\frac{|\alpha| + D}{2}}
  \exp \left( -\frac{|\bv|^2}{2} \right)
  \prod_{d=1}^D \He_{\alpha_d}(v_d),
\end{equation}
where $\He_n(x)$'s are the orthogonal Hermite polynomials defined by
\begin{equation}
\He_n(x) = (-1)^n \frac{\mathrm{d}^n}{\mathrm{d}x^n} \exp(-x^2/2).
\end{equation}
This idea originates from \cite{Grad} where the 13-moment equations
are derived. In order to get a finite system, the \NRxx method
\cite{NRxx_new} chooses an positive integer $M$ and approximate
$f_{\alpha}$ with $|\alpha| = M + 1$ by
\begin{equation} \label{eq:regularization}
f_{\alpha} = -\frac{\theta}{\nu} \sum_{j=1}^D
  \frac{\partial f_{\alpha - e_j}}{\partial x_j},
\qquad |\alpha| = M + 1.
\end{equation}
Now that all $f_{\alpha}$'s with $|\alpha| = M + 1$ have been related
with $f_{\beta}$'s with $|\beta| = M$, we can get a closed moment
system of all $f_{\gamma}$'s with $|\gamma| \leqslant M$ by putting
the expansion \eqref{eq:expansion} into the BGK equation
\eqref{eq:BGK}.

The explicit expressions of such moment systems can be written in a
uniform style for any choice of $M$, but those expressions are not
convenient for computation, since they are not in a conservative
form. In order to avoid an intricate process to obtain balance laws,
the \NRxx method treats the distribution function \eqref{eq:expansion}
as a whole, instead of considering each moment $f_{\alpha}$ as an
individual variable. For the construction of an applicable scheme, it
is necessary to provide a method which is able to apply addition or
subtraction on the following two distributions:
\begin{equation}
f_1(\bxi) = \sum_{\alpha \in \bbN^D} f_{1,\alpha}
  \mathcal{H}_{\theta_1, \alpha} \left(
    \frac{\bxi - \bu_1}{\sqrt{\theta_1}}
  \right), \qquad
f_2(\bxi) = \sum_{\alpha \in \bbN^D} f_{2,\alpha}
  \mathcal{H}_{\theta_2, \alpha} \left(
    \frac{\bxi - \bu_2}{\sqrt{\theta_2}}
  \right).
\end{equation}
Additionally, due to the cutoff, only $f_{1,\alpha}$'s and
$f_{2,\alpha}$'s with $|\alpha| \leqslant M + 1$ are known. In
\cite{NRxx}, the authors proposed a homotopic method to calculate a
new representation of $f_1$:
\begin{equation} \label{eq:projected}
f_1(\bxi) = \sum_{\alpha \in \bbN^D} \tilde{f}_{1,\alpha}
  \mathcal{H}_{\theta_2, \alpha} \left(
    \frac{\bxi - \bu_2}{\sqrt{\theta_2}}
  \right),
\end{equation}
where the coefficients $\tilde{f}_{1,\alpha}$'s with $|\alpha|
\leqslant M + 1$ can be worked out by solving the ODE system
\begin{equation} \label{eq:ode}
\left\{ \begin{array}{l}
\displaystyle \frac{\mathrm{d}}{\mathrm{d} \tau} F_{\alpha} =
  [1 - \tau R(\tau)]^2 \sum_{d=1}^D \left[
    R(\tau) \theta_1 F_{\alpha - 2e_d}
    + (u_{1,d} - u_{2,d}) \sqrt{\theta_1 / \theta_2}
      F_{\alpha - e_d}
  \right], \\
F_{\alpha}(0) = f_{1,\alpha}
\end{array} \right.
\end{equation}
for all $|\alpha| \leqslant M + 1$ until $\tau = 1$, and setting
$\tilde{f}_{1,\alpha} = F_{\alpha}(1)$. In Eq. \eqref{eq:ode},
$R(\tau)$ is defined as
\begin{equation}
R(\tau) = \frac{\sqrt{\theta_1} - \sqrt{\theta_2}}
  {(\sqrt{\theta_1} - \sqrt{\theta_2}) \tau - \sqrt{\theta_2}}.
\end{equation}
The system \eqref{eq:ode} will be solved by a Runge-Kutta type scheme.
Once the form \eqref{eq:projected} is obtained, $f_1 + f_2$ or $f_1 -
f_2$ can be calculated naturally.

By solving \eqref{eq:ode}, we are able to represent $f$ for any $\bu'$
and $\theta'$ as
\begin{equation} \label{eq:representation}
f(\bxi) = \sum_{|\alpha| \leqslant M + 1} f_{\alpha}'
  \mathcal{H}_{\theta', \alpha} \left(
    \frac{\bxi - \bu'}{\sqrt{\theta'}}
  \right) + \cdots,
\end{equation}
once there is one such representation known for some particular $\bu'$
and $\theta'$. Here the ellipsis means the remaining coefficients are
unknown. For any $\bu'$ and $\theta'$ and the associated
representation of $f$ \eqref{eq:representation}, we have
\begin{equation} \label{eq:macro_vars}
\begin{aligned}
\rho &= \int_{\bbR^D} f(\bxi) \dd \bxi = f_0', \\
\rho u_d &= \int_{\bbR^D} \xi_d f(\bxi) \dd \bxi
  = f_0' u_d' + f_{e_d}', \quad d = 1,\cdots,D, \\
\rho |\bu|^2 + D\rho \theta &
  = \int_{\bbR^D} |\bxi|^2 f(\bxi) \dd \bxi
  = \rho |\bu'|^2 +
    \sum_{d=1}^D (\theta' f_0' + 2 f_{e_d}' u_d' + 2 f_{2e_d}').
\end{aligned}
\end{equation}
When $\bu' = \bu$ and $\theta' = \theta$, \eqref{eq:representation} is
called as the standard representation of $f$. Note that the
coefficients in \eqref{eq:expansion} and \eqref{eq:regularization} are
in the sense of standard representation.

Another important technique is to calculate the flux $\bxi f$
based on the Hermite expansion of $f$. Since only the moments with
$|\alpha| \leqslant M + 1$ are known, $\bxi f$ can only be accurately
given upto the $M$-th order moments. Suppose $f$ is presented as
\eqref{eq:representation} for some $\bu'$ and $\theta'$, and we let
\begin{equation}
F_j(\bxi) = \xi_j f(\bxi) = \sum_{|\alpha| \leqslant M}
  F_{j,\alpha}' \mathcal{H}_{\theta', \alpha} \left(
    \frac{\bxi - \bu'}{\sqrt{\theta'}}
  \right) + \cdots, \quad j = 1,\cdots,D.
\end{equation}
Then, by making use of the recurrence relation of the Hermite
polynomials, we have
\begin{equation}
F_{j,\alpha}' = \theta' f_{\alpha-e_j}' + u_j' f_{\alpha}'
  + (\alpha_j + 1) f_{\alpha + e_j}', \quad |\alpha| \leqslant M.
\end{equation}
Here $f_{\alpha - e_j}'$ is taken to be zero if $\alpha = 0$.

Once the linear operations between discrete distributions and the
calculation of the flux are applicable, the convection progress can be
simulated by a Riemann-solver-free finite volume method, such as
Lax-Friedrichs scheme and HLL scheme (see e.g. \cite {FVM}). Suppose
the problem is one-dimensional, and the spatial mesh is uniform with
mesh size $\Delta x$. We denote by $f_i^n$ the distribution function
on the $i$-th grid at the time step $t_n$. Other symbols such as
$\bu_i^n$ and $\theta_i^n$ are defined similarly.  Then, a first-order
scheme with HLL numerical flux can be described as follows:
\begin{enumerate}
\item Let $n = 0$ and set $f_i^n(\bxi)$ to be the initial value, which
  is in its standard representation.
\item \label{item:diffusion} Apply \eqref{eq:regularization} to obtain
  the $(M+1)$-st order moments:
  \begin{equation}
  f_{i,\alpha}^n = -\frac{\theta_i^n}{\nu_i^n}
    \frac{f_{i+1,\alpha-e_1}^n - f_{i-1,\alpha-e_1}^n}{2\Delta x},
  \quad |\alpha| = M + 1.
  \end{equation}
\item \label{item:convection} Solve the convection part with the HLL
  scheme:
  \begin{equation} \label{eq:FVM}
  f_i^{n*}(\bxi) = f_i^n(\bxi) - \frac{\Delta t}{\Delta x}
    [G_{i+1/2}^n(\bxi) - G_{i-1/2}^n(\bxi)],
  \end{equation}
  where
  \begin{equation} \label{eq:flux}
  G_{i+1/2}^n(\bxi) = \left\{ \begin{array}{ll}
    \xi_1 f_i^n(\bxi), & 0 \leqslant \lambda_{i+1/2}^L, \\[5pt]
    \dfrac{\lambda_{i+1/2}^R \xi_1 f_i^n(\bxi) -
      \lambda_{i+1/2}^L \xi_1 f_{i+1}^n(\bxi)}
      {\lambda_{i+1/2}^R - \lambda_{i+1/2}^L} \\[15pt]
    \qquad + \dfrac{\lambda_{i+1/2}^L \lambda_{i+1/2}^R
        [f_{i+1}^n(\bxi) - f_i^n(\bxi)]}
      {\lambda_{i+1/2}^R - \lambda_{i+1/2}^L}, &
      \lambda_{i+1/2}^L < 0 < \lambda_{i+1/2}^R, \\[15pt]
    \xi_1 f_{i+1}^n(\bxi), & 0 \geqslant \lambda_{i+1/2}^R.
  \end{array} \right.
  \end{equation}
  In Eq. \eqref{eq:flux}, the signal speed $\lambda_{i+1/2}^L$ and
  $\lambda_{i+1/2}^R$ are approximated by
  \begin{equation} \label{eq:sig_speed}
  \begin{aligned}
  \lambda_{i+1/2}^L &= \min\left\{
    u_{1,i}^n - C_{M+1} \sqrt{\theta_i^n},
    \: u_{1,i+1}^n - C_{M+1} \sqrt{\theta_{i+1}^n}
  \right\}, \\
  \lambda_{i+1/2}^R &= \max\left\{
    u_{1,i}^n + C_{M+1} \sqrt{\theta_i^n},
    \: u_{1,i+1}^n + C_{M+1} \sqrt{\theta_{i+1}^n}
  \right\},
  \end{aligned}
  \end{equation}
  where $C_{M+1}$ is the maximal root of Hermite polynomial
  $\He_{M+1}(x)$. Eq. \eqref{eq:sig_speed} is also used to determine
  the time step $\Delta t$. We refer to \cite{NRxx} for details. Note
  that \eqref{eq:FVM} is only accurate up to the $M$th order moments.
\item \label{item:collision} Solve the collision part analytically:
  \begin{enumerate}
  \item For each $f_i^{n*}(\bxi)$, get its standard representation
    using \eqref{eq:ode} and \eqref{eq:macro_vars}. We suppose the
    result is
    \begin{equation}
    f_i^{n*}(\bxi) = \sum_{|\alpha| \leqslant M} f_{i,\alpha}^{n*}
      \mathcal{H}_{\theta_i^{n+1}, \alpha} \left(
        \frac{\bxi - \bu_i^{n+1}}{\sqrt{\theta_i^{n+1}}}
      \right) + \cdots.
    \end{equation}
  \item Multiply all coefficients $f_{i,\alpha}^{n*}$, $|\alpha|
    \geqslant 2$ by $\exp(-\nu \Delta t)$. The result is
    \begin{equation} \label{eq:new_sol}
    f_i^{n+1}(\bxi) = \sum_{|\alpha| \leqslant M} f_{i,\alpha}^{n+1}
      \mathcal{H}_{\theta_i^{n+1}, \alpha} \left(
        \frac{\bxi - \bu_i^{n+1}}{\sqrt{\theta_i^{n+1}}}
      \right) + \cdots,
    \end{equation}
    where 
    \begin{equation} \label{eq:collision}
    f_{i,\alpha}^{n+1} = \left\{ \begin{array}{ll}
      f_{i,\alpha}^{n*}, & |\alpha| < 2, \\[5pt]
      f_{i,\alpha}^{n*} \exp(-\nu_i^n \Delta t), & |\alpha| \geqslant 2.
    \end{array} \right.
    \end{equation}
  \end{enumerate}
\item Increase $n$ by $1$ and return to step \ref{item:diffusion}.
\end{enumerate}
In step \ref{item:convection}, since diffusion terms exist in the
equations, the time step satisfies $\Delta t = O(\Delta x^2)$. The
whole scheme is of first order.


\section{A high resolution scheme} \label{sec:reconstruction} As is
well known, the first-order HLL flux \eqref{eq:flux} adds excessive
diffusion to the numerical solution in a general case. In order to
reduce numerical diffusion, the technique of linear reconstruction is
introduced below to the finite volume method. Suppose the boundary
between the $i$-th and $(i+1)$-st cells is located at $x =
x_{i+1/2}$. Our aim is to construct two distributions
$f_{i+1/2}^L(\bxi)$ and $f_{i+1/2}^R(\bxi)$ based on all cell averages
$f_i$ to approximate the left and right limit of $f(x, \bxi)$ at point
$x = x_{i+1/2}$. In this section, the superscript $n$ will be omitted
since the reconstruction is only applied on the same time step. Thus
the numerical flux in \eqref{eq:FVM} can be re-formulated as
\begin{equation} \label{eq:new_flux}
G_{i+1/2}(\bxi) = \left\{ \begin{array}{ll}
  \xi_1 f_{i+1/2}^L(\bxi), &
    0 \leqslant \lambda_{i+1/2}^L, \\[5pt]
  \dfrac{\lambda_{i+1/2}^R \xi_1 f_{i+1/2}^L(\bxi) -
    \lambda_{i+1/2}^L \xi_1 f_{i+1/2}^R(\bxi)}
    {\lambda_{i+1/2}^R - \lambda_{i+1/2}^L} \\[15pt]
  \qquad + \dfrac{\lambda_{i+1/2}^L \lambda_{i+1/2}^R
      [f_{i+1/2}^R(\bxi) - f_{i+1/2}^L(\bxi)]}
    {\lambda_{i+1/2}^R - \lambda_{i+1/2}^L}, &
    \lambda_{i+1/2}^L < 0 < \lambda_{i+1/2}^R, \\[15pt]
  \xi_1 f_{i+1/2}^R(\bxi), & 0 \geqslant \lambda_{i+1/2}^R,
\end{array} \right.
\end{equation}
where
\begin{equation}
\begin{aligned}
\lambda_{i+1/2}^L &= \min\left\{
  u_{1,i+1/2}^L - C_{M+1} \sqrt{\theta_{i+1/2}^L},
  \: u_{1,i+1/2}^R - C_{M+1} \sqrt{\theta_{i+1/2}^R}
\right\}, \\
\lambda_{i+1/2}^R &= \max\left\{
  u_{1,i+1/2}^L + C_{M+1} \sqrt{\theta_{i+1/2}^L},
  \: u_{1,i+1/2}^R + C_{M+1} \sqrt{\theta_{i+1/2}^R}
\right\}.
\end{aligned}
\end{equation}

During the reconstruction, different methods are applied to
the convection part ($|\alpha| \leqslant M$) and the diffusion part
($|\alpha| = M + 1$).

\subsection{Reconstruction for the convection part}
For the convection part, it is important that the quantities used in
reconstruction are conservative variables. Before reconstruction,
according to the algorithm described in section \ref{sec:NRxx}, we
already have the standard representations for all distributions
$f_i(\bxi)$, and the coefficients are assumed to be $f_{i,\alpha}$,
$|\alpha| \leqslant M$. The simplest idea is to use $f_{i,\alpha}$
together with $\bu_i$ and $\theta_i$ to make linear reconstruction:
\begin{equation} \label{eq:wrong_reconstruction}
\begin{gathered}
f_{i-1/2,\alpha}^R = f_{i,\alpha} - g_{i,\alpha} \Delta x / 2, \quad
f_{i+1/2,\alpha}^L = f_{i,\alpha} + g_{i,\alpha} \Delta x / 2, \\
\bu_{i-1/2}^R = \bu_i - \boldsymbol{g}_i \Delta x / 2, \quad
\bu_{i+1/2}^L = \bu_i + \boldsymbol{g}_i \Delta x / 2, \\
\theta_{i-1/2}^R = \theta_i - g_i \Delta x / 2, \quad
\theta_{i+1/2}^L = \theta_i + g_i \Delta x / 2,
\end{gathered}
\end{equation}
and
\begin{equation} \label{eq:std_representation}
\begin{gathered}
f_{i-1/2}^R(\bxi) = \sum_{|\alpha| \leqslant M}
  f_{i-1/2,\alpha}^R \mathcal{H}_{\theta_{i-1/2}^R, \alpha}
  \left(
    \frac{\bxi - \bu_{i-1/2}^R}{\sqrt{\theta_{i-1/2}^R}}
  \right) + \cdots, \\
f_{i+1/2}^L(\bxi) = \sum_{|\alpha| \leqslant M}
  f_{i+1/2,\alpha}^L \mathcal{H}_{\theta_{i+1/2}^L, \alpha}
  \left(
    \frac{\bxi - \bu_{i+1/2}^L}{\sqrt{\theta_{i+1/2}^L}}
  \right) + \cdots,
\end{gathered}
\end{equation}
where $g_{i,\alpha}$ and $g_i$ are constants, and $\boldsymbol{g}_i$
is a constant vector. However, this method leads to incorrect
numerical results since none of the variables in Eq. \eqref
{eq:wrong_reconstruction} is conservative.

We take the original idea of the \NRxx method, and consider the
distribution function as a whole. Thus, linear reconstruction means
taking the following approximation of $f_{i-1/2}^R(\bxi)$ and
$f_{i+1/2}^L(\bxi)$:
\begin{equation} \label{eq:reconstruction}
f_{i-1/2}^R(\bxi) = f_i(\bxi) - \frac{\Delta x}{2} g_i(\bxi), \quad
f_{i+1/2}^L(\bxi) = f_i(\bxi) + \frac{\Delta x}{2} g_i(\bxi).
\end{equation}
Here $g_i(\bxi)$ is a distribution. Obviously, the most convenient
representation of $g_i(\bxi)$ is
\begin{equation} \label{eq:g_i}
g_i(\bxi) = \sum_{|\alpha| \leqslant M} g_{i,\alpha}
  \mathcal{H}_{\theta_i, \alpha}
  \left( \frac{\bxi - \bu_i}{\sqrt{\theta_i}} \right) + \cdots.
\end{equation}
Thus no ODEs are to be solved during the calculation of \eqref
{eq:reconstruction}. Now, the coefficients $g_{i,\alpha}$'s can be
naturally given if we represent $f_{i-1}(\bxi)$ and $f_{i+1}(\bxi)$ as
\begin{equation}
\begin{gathered}
f_{i-1}(\bxi) = \sum_{|\alpha| \leqslant M} f_{i-1,\alpha}^i
  \mathcal{H}_{\theta_i, \alpha}
  \left( \frac{\bxi - \bu_i}{\sqrt{\theta_i}} \right) + \cdots, \\
f_{i+1}(\bxi) = \sum_{|\alpha| \leqslant M} f_{i+1,\alpha}^i
  \mathcal{H}_{\theta_i, \alpha}
  \left( \frac{\bxi - \bu_i}{\sqrt{\theta_i}} \right) + \cdots.
\end{gathered}
\end{equation}
In the implementation, we use the simplest minmod slope limiter for
reconstruction
\begin{equation} \label{eq:g_i_coef}
g_{i,\alpha} = \mathrm{minmod} \left\{
  \frac{f_{i+1,\alpha}^i - f_{i,\alpha}}{\Delta x},
  \frac{f_{i,\alpha} - f_{i-1,\alpha}^i}{\Delta x}
\right\}.
\end{equation}
This reconstruction is a conservative reconstruction, since
\begin{equation}
\int_{x_{i-1/2}}^{x_{i+1/2}} \tilde{f}_i(x, \bxi)
  = \Delta x f_i(\bxi), \quad \forall i \in \mathbb{Z},
\end{equation}
where $\tilde{f}_i(x,\bxi)$ is a linear function of $x$ defined on
$(x_{i-1/2}, x_{i+1/2})$ as
\begin{equation}
\tilde{f}_i(x,\bxi) = f_i(\bxi) + g_i(\bxi)
  \left( x - \frac{x_{i-1/2} + x_{i+1/2}}{2} \right),
\quad x \in (x_{i-1/2}, x_{i+1/2}).
\end{equation}
However, this condition is not satisfied by the reconstruction
\eqref{eq:wrong_reconstruction}.

\subsection{Reconstruction for the diffusion part}
The diffusion terms \eqref{eq:regularization} provide approximation to
the $(M+1)$-st order moments. Since \eqref{eq:regularization} is in the
sense of standard representation, we first need to get the standard
representations of $f_{i-1/2}^R(\bxi)$ and $f_{i+1/2}^L(\bxi)$, and we
use \eqref{eq:std_representation} to denote the results. The
reconstruction of $f_{i-1/2,\alpha}^R$ and $f_{i+1/2,\alpha}^L$ with
$|\alpha| = M+1$, which are involved in the ellipsis of Eq. \eqref
{eq:std_representation}, is a direct discretization of \eqref
{eq:regularization}:
\begin{equation} \label{eq:diffusion}
\begin{array}{l}
f_{i-1/2,\alpha}^R = -\dfrac{\theta_{i-1/2}^R}{\nu_{i-1/2}^R}
  \dfrac{f_{i, \alpha - e_1} - f_{i-1, \alpha - e_1}}{\Delta x}, \\
f_{i+1/2,\alpha}^L = -\dfrac{\theta_{i+1/2}^L}{\nu_{i+1/2}^L}
  \dfrac{f_{i+1, \alpha - e_1} - f_{i, \alpha - e_1}}{\Delta x},
\end{array} \qquad |\alpha| = M + 1.
\end{equation}

Now we give a general discussion on the order of accuracy. The time
splitting introduces an error of magnitude $O(\Delta t) = O(\Delta
x^2)$. With \eqref{eq:reconstruction}, \eqref{eq:g_i} and
\eqref{eq:g_i_coef}, the numerical flux \eqref{eq:new_flux} turns out
to be a second order numerical flux. However, as discussed in
\cite{Torrilhon2006, Xu}, due to the one-sided approximation of the
diffusive gradients \eqref{eq:diffusion}, the final accuracy only
appears to be the first order. However, comparing with the original
scheme described in Section \ref{sec:NRxx}, numerical error is
significantly reduced by the linear reconstruction.


\section{Enlarging the time step} \label{sec:time_step}
The regularization of the moment method introduces diffusion terms
into the system, which yields a relatively small time step $\Delta t =
O(\Delta x^2)$. In order to enlarge the time step length, we use the
RKC time stepping in the temporal discretization. In this section, a
large time-stepping scheme with 2nd-order time integration will be
proposed.

\subsection{The RKC time-stepping} \label{sec:RKC}
The RKC method is a series of explicit Runge-Kutta schemes for
parabolic problems with large stability region and good internal
stability. Our aim is to use $\Delta t = O(\Delta x)$ in our algorithm
without producing errors larger than the original method.  Thus a
second-order RKC scheme is needed. The $s$-stage second-order RKC
formula for the ODE system
\begin{equation}
w'(t) = F(w(t))
\end{equation}
was deduced in \cite{RKC} as
\begin{equation}
\begin{aligned}
& W_0 = w^n, \\
& W_1 = W_0 + \tilde{\mu}_1 \Delta t F_0, \\
& W_j = (1 - \mu_j - \nu_j) W_0 + \mu_j W_{j-1} + \mu_j W_{j-2}
  + \tilde{\mu}_j \Delta t F_{j-1} + \tilde{\gamma}_j \Delta t F_0,
  \quad j = 2, \cdots, s, \\
& w^{n+1} = W_s,
\end{aligned}
\end{equation}
where $F_k = F(W_k)$, $\tilde{\mu}_1 = b_1 \omega_1$, and
\begin{equation}
\mu_j = \frac{2 b_j \omega_0}{b_{j-1}}, \quad
\nu_j = -\frac{b_j}{b_{j-2}}, \quad
\tilde{\mu}_j = \frac{2 b_j \omega_1}{b_{j-1}}, \quad
\tilde{\gamma}_j = -a_{j-1} \tilde{\mu}_j, \qquad
j = 2,\cdots,s.
\end{equation}
Here the parameters $a_j$, $b_j$, $\omega_0$, $\omega_1$ are relevant
to a manually selected damping factor $\epsilon$. They are given by
\begin{equation}
\begin{aligned}
& \omega_0 = 1 + \epsilon s^2, \quad
  \omega_1 = T_s'(\omega_0) / T_s''(\omega_0), \\
& b_0 = b_1 = b_2, \quad b_j = T_j''(\omega_0) / (T_j'(\omega_0))^2,
  \qquad j = 2, \cdots, s, \\
& a_j = 1 - b_j T_j(\omega_0), \qquad j = 1, \cdots, s-1,
\end{aligned}
\end{equation}
where $T_j(x)$ is the first kind Chebyshev polynomials
\begin{equation}
T_j(x) = \cos(j \arccos x) = \cosh (j \mathop{\mathrm{arccosh}} x),
  \qquad j = 0,1,2,\cdots.
\end{equation}
which can also be defined by the recurrence relation
\begin{equation}
T_0(x) = 1, \quad T_1(x) = x, \quad
T_{j+1}(x) = 2x T_j(x) - T_{j-1}(x), \qquad j = 1,2,\cdots.
\end{equation}
The damping factor $\epsilon$ is often chosen as a small positive
value such that a small imaginary perturbation of the eigenvalues of
$F'(w)$ is allowable. In \cite{Hundsdorfer}, the authors suggest that
$\epsilon$ be chosen as $2/13$, which results in a reduction in the
stability boundary of about 2\%.

For advection-diffusion problems, the imaginary part of the
eigenvalues of $F'(w)$ may be large. Some analysis of the
eigenvalue structure of upwind finite volume methods can be found in
\cite{Jeltsch}, where the authors show that the imaginary parts of the
eigenvalues are less than $2.0$ for the second-order scheme if the
time step satisfies the CFL condition of a pure advection problem. In
order to ensure the stability, we follow \cite{Verwer} and use a large
damping factor $\epsilon = 10$. Thus the stability boundary is
approximately given by
\begin{equation}
\beta(s) \approx 0.34 (s^2 - 1).
\end{equation}

In our implementation, the RKC method is only applied to the finite
volume scheme \eqref{eq:FVM}. The time step is determined by
\begin{equation} \label{eq:CFL}
\frac{\lambda_{\max} \Delta t}{\Delta x} \leqslant \mathit{CFL},
\end{equation}
where $\lambda_{\max}$ is defined as
\begin{equation}
\lambda_{\max} = \max_i \left\{
  \lambda_{i+1/2}^L, \lambda_{i+1/2}^R
\right\}.
\end{equation}
Then, we find the smallest positive integer $s$ satisfying
\begin{equation} \label{eq:RKC_stage}
\Delta t \left( \frac{\lambda_{\max}}{\Delta x}
  + \frac{2(M+1)}{(\Delta x)^2} \left(
    \frac{\theta}{\nu}
  \right)_{\max} \right)
  \leqslant \frac{1}{2} \mathit{CFL} \cdot \beta(s),
\end{equation}
and use the $s$-stage RKC method instead of the forward Euler scheme
in \eqref{eq:FVM}. The inequalities \eqref{eq:CFL} and
\eqref{eq:RKC_stage} give $\Delta t = O(\Delta x)$ and $s =
O\left(1/\sqrt{\Delta x}\right)$, which leads to the following
estimation of the total computational time:
\begin{equation}
t_{\mathrm{com}} \approx \frac{T}{\Delta t / s}
  \cdot \frac{L}{\Delta x} = O(\Delta x^{-2.5}),
\end{equation}
where $L$ is the length of the computational domain, and $T$ is the
finishing time of the computation.

\subsection{The Strang splitting}
When the RKC time stepping is used, according to \eqref{eq:CFL}, the
time step has the same magnitude as the grid size. Thus, when the gas
is dense, we can still have only the first-order accuracy due to the
convection-collision splitting, which introduces an error $O(\Delta
t) = O(\Delta x)$. In order to restore the time integration to a
second-order one, the Strang splitting technique \cite{Strang} is
employed. In our algorithm, a direct usage of the Strang splitting can
be described as follows:
\begin{enumerate}
\setlength\itemsep{0cm}
\item Let $n = 0$.
\item \label{item:dt} Determine the time step $\Delta t_n$.
\item \label{item:first_collision} Solve the collision part over a
  half time step of length $\Delta t_n / 2$.
\item Solve the convection part using the second-order RKC scheme over
  a time step of length $\Delta t_n$.
\item \label{item:second_collision} Solve the collision part again
  over a half time step of length $\Delta t_n / 2$.
\item Increase $n$ by $1$ and return to Step \ref{item:dt}.
\end{enumerate}
This scheme requires twice calculation of the collision operator in
one time step, but decrease the time integration error to $O(\Delta
t^2)$. Here we introduce a method equivalent to the above Strang
splitting scheme, but only once calculation of collision term is
needed per time step.

According to \eqref{eq:CFL} and \eqref{eq:sig_speed}, the time step
determined in Step \ref{item:dt} is only relevant to $\bu$ and
$\theta$. Since the collision steps (Step \ref{item:first_collision}
and Step \ref{item:second_collision}) do not change these two
quantities, we can actually calculate $\Delta t_{n+1}$ before Step
\ref{item:second_collision}. Thus the above algorithm can be
rearranged as follows:
\begin{enumerate}
\setlength\itemsep{0cm}
\item Let $n = 0$ and determine the time step $\Delta t_0$.
\item Solve the collision part over a half time step of length $\Delta
  t_n / 2$.
\item \label{item:conv} Solve the convection part using the
  second-order RKC scheme over a time step of length $\Delta t_n$.
\item Determine the next time step $\Delta t_{n+1}$.
\item \label{item:coll} Solve the collision part over a half time step
  of length $\Delta t_n / 2$, and then solve the collision part again
  over a half time step of length $\Delta t_{n+1} / 2$.
\item Increase $n$ by $1$ and return to Step \ref{item:conv}.
\end{enumerate}
Recall that for the BGK model, the collision part can be solved
analytically as \eqref{eq:new_sol} and \eqref{eq:collision}.
Therefore, we can replace Step \ref{item:coll} by
\begin{enumerate}
\setcounter{enumi}{4}
\item Solve the collision part over a time step of length $(\Delta t_n
  + \Delta t_{n+1}) / 2$.
\end{enumerate}
Thus, the numerical result is identical to the original Strang
splitting scheme, but only one collision step is performed in a
complete time step.


\section{Numerical examples} \label{sec:num_ex} In this section, two
numerical examples are presented to validate the efficiency and
accuracy of our algorithm. In all the tests, the collision frequency
$\nu$ is given by a simple form $\rho / \Kn$, which is corresponding
to the Maxwell molecules. Here $\Kn$ is the global Knudsen number, and
it is slightly different from the Knudsen number defined in
\cite{Bird} (denoted by $\Kn'$) by
\begin{equation} \label{eq:Knudsen}
\Kn' = \frac{8}{5} \sqrt{\frac{2}{\pi}} Kn.
\end{equation}
The CFL number is chosen to be $0.95$. All the computations are
performed on the Dell OptiPlex 755 desktop computer with a dual-core
processor and CPU speed 2.33GHz.

\subsection{An example with smooth solution} \label{sec:periodic}
\begin{figure}[p]
\centering
\subfigure[$M=3$, density]{
  \includegraphics[width=.45\textwidth]{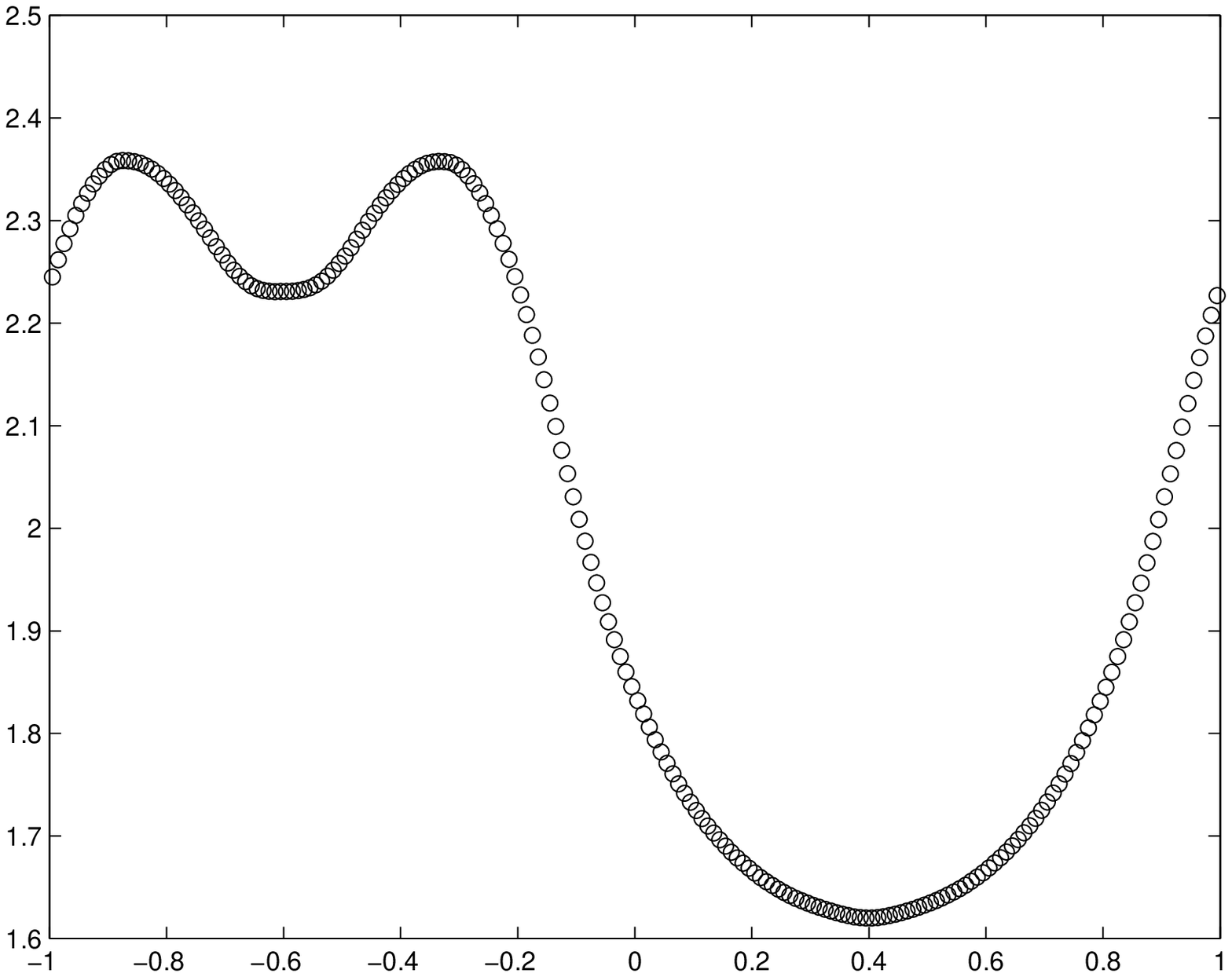}
}
\subfigure[$M=3$, temperature]{
  \includegraphics[width=.45\textwidth]{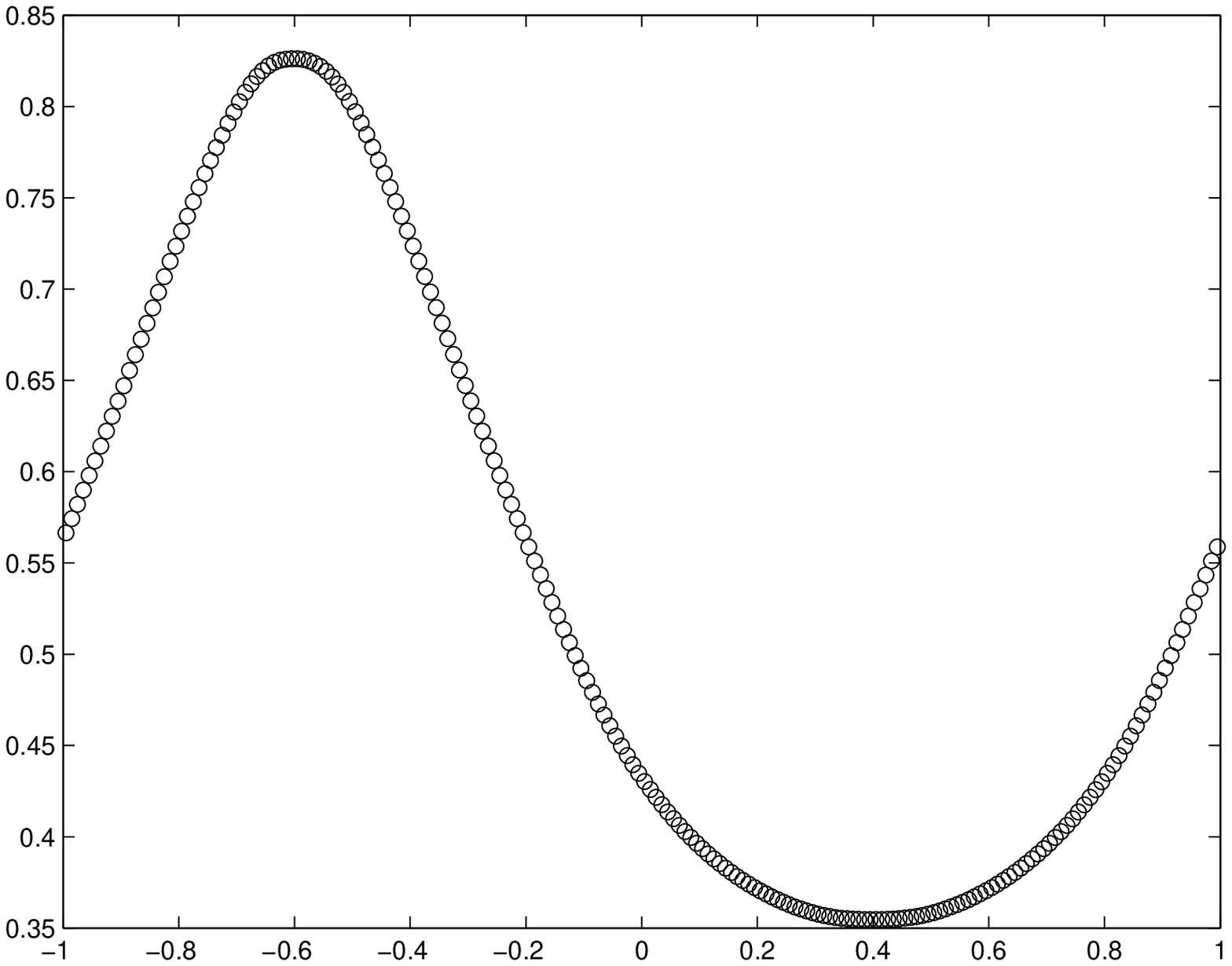}
}
\subfigure[$M=6$, density]{
  \includegraphics[width=.45\textwidth]{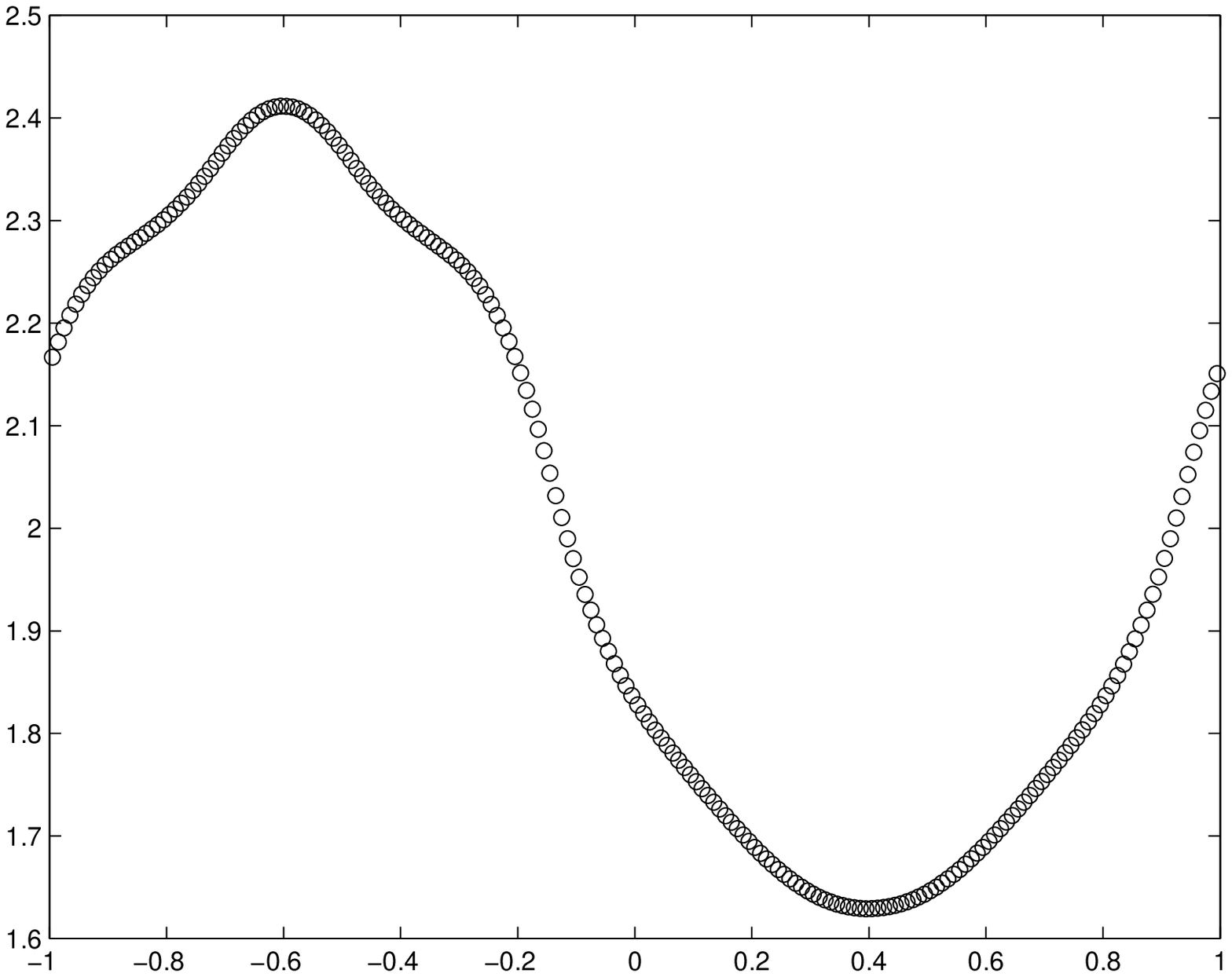}
}
\subfigure[$M=6$, temperature]{
  \includegraphics[width=.45\textwidth]{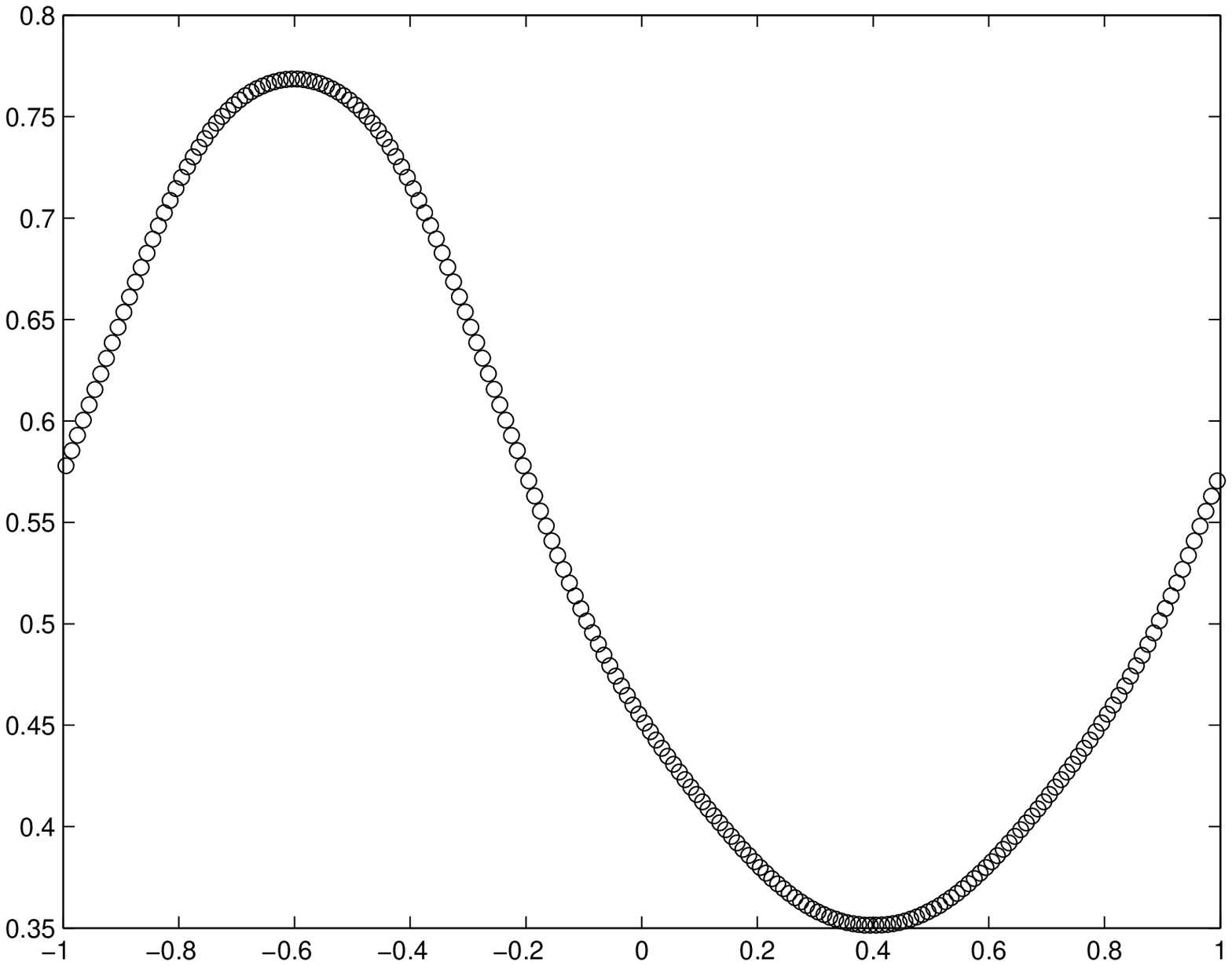}
}
\subfigure[$M=9$, density]{
  \includegraphics[width=.45\textwidth]{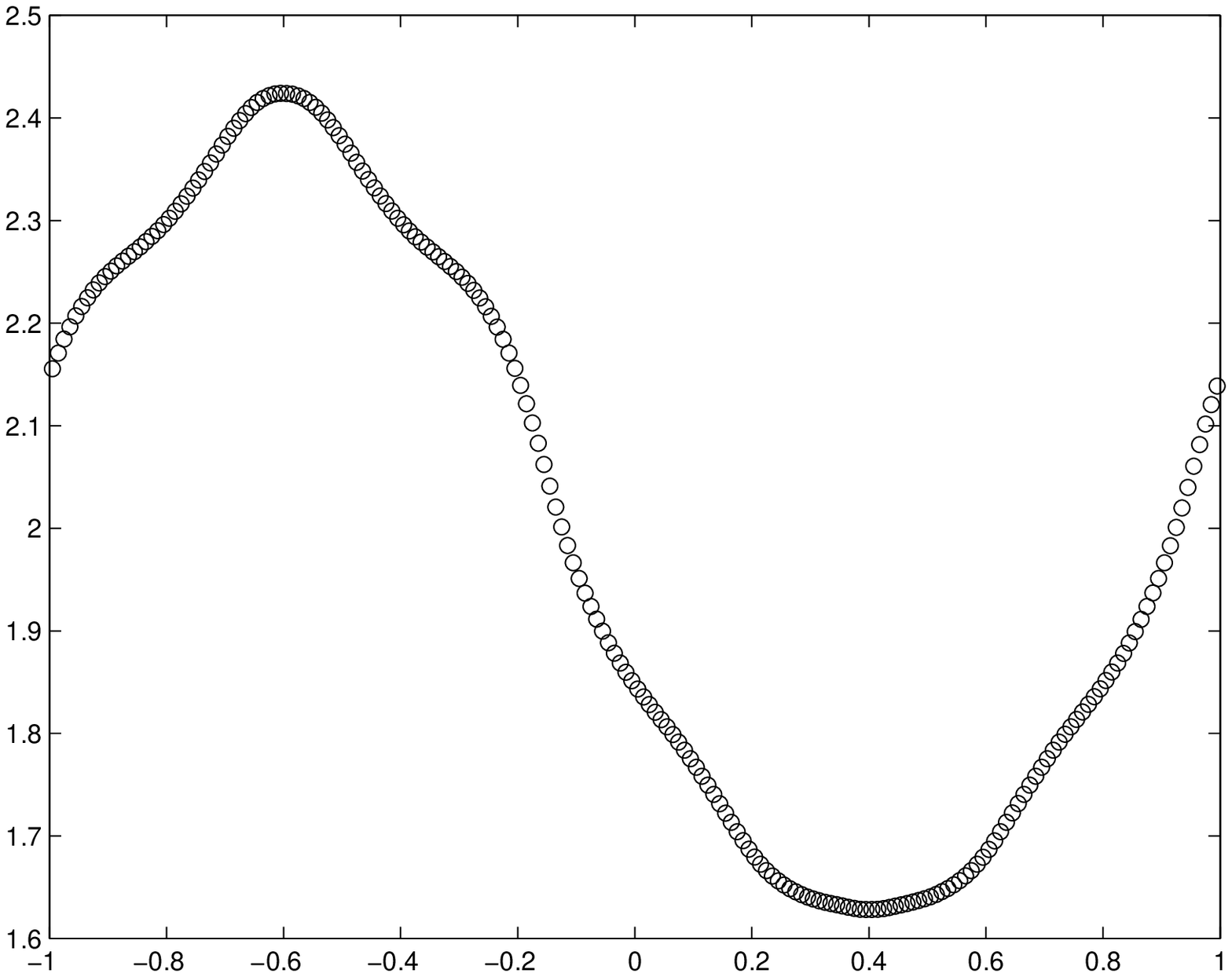}
}
\subfigure[$M=9$, temperature]{
  \includegraphics[width=.45\textwidth]{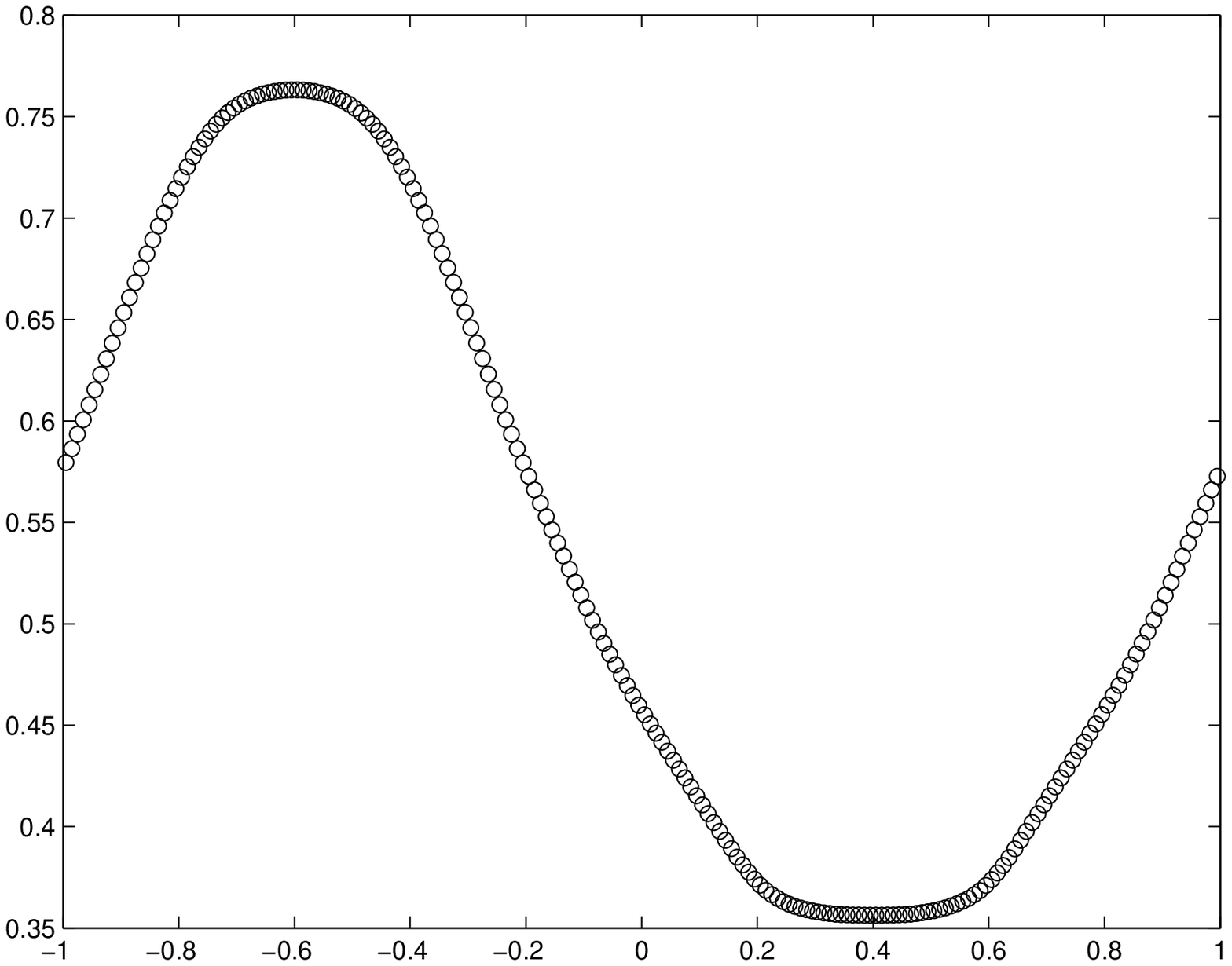}
}
\caption{Numerical solutions of Problem \ref{sec:periodic} using $200$
spatial grids}
\label{fig:periodic:solution}
\end{figure}

The first example is a repetition of the 1D periodic problem in
\cite{Torrilhon2006,NRxx}. The computational domain is $[-1, 1]$ and
the boundary condition is assumed to be periodic. The initial
condition is
\begin{equation}
\rho_0(x) = 2 + \frac{1}{2} \cos(\pi x), \quad
\bu_0(x) = \left(
  1 + \frac{1}{2} \sin(\pi x), \frac{1}{2} \sin(\pi x), 0
\right)^T, \quad
p_0(x) = 1,
\end{equation}
and the distribution is the local Maxwellian everywhere. The
computation is stopped at $t = 0.4$.  We set $\Kn = 0.5$ and $M =
3,6,9$. The numerical results for the density $\rho$ and temperature
$\theta$ are plotted in Figure \ref{fig:periodic:solution}. Since the
Knudsen number is large, the profiles of density and temperature
differ a lot for different moment systems.

\begin{figure}[p]
\centering
\subfigure[Time cost $t_{\mathrm{com}}$]{
  \includegraphics[width=.5\textwidth]{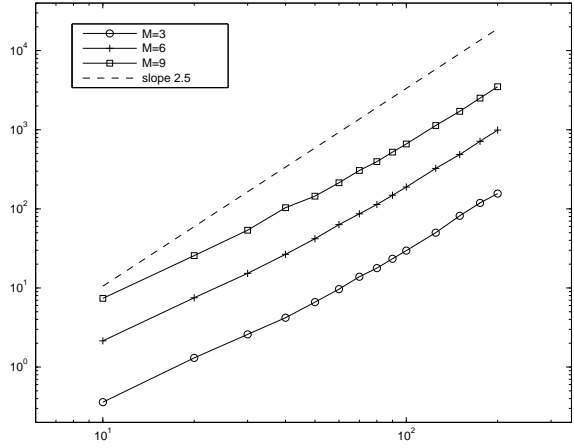}
}
\subfigure[Average step size $\Delta t$]{
  \includegraphics[width=.5\textwidth]{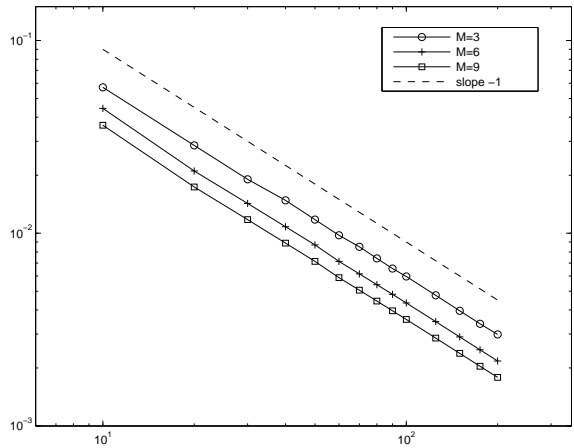}
}
\subfigure[Average effective step size $\Delta t / s$]{
  \includegraphics[width=.5\textwidth]{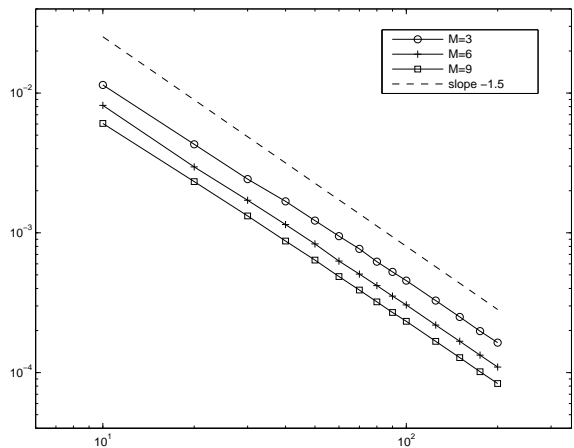}
}
\caption{Computational costs and time step sizes for Problem
\ref{sec:periodic} on different spatial grids. The $x$-axis is the
logarithm of the grid number $N$.}
\label{fig:periodic:efficiency}
\end{figure}

In order to examine the efficiency of our algorithm, we discretize the
problem on a series of spatial grids with grid numbers ranging from
$10$ to $200$. Using $N = 2/\Delta x$, the analysis at the end of
Section \ref{sec:RKC} gives
\begin{equation}
t_{\mathrm{com}} = O(N^{2.5}), \quad
\Delta t = O(1 / N), \quad \Delta t / s = O(1 / N^{1.5}).
\end{equation}
These results are validated by Figure \ref{fig:periodic:efficiency}.
We can see that the large time step sizes are achieved, and the
results are still stable. This leads to a remarkable reduction of the
computational time.

\begin{figure}[p]
\centering
\subfigure[$M=3$]{
  \includegraphics[width=.5\textwidth]{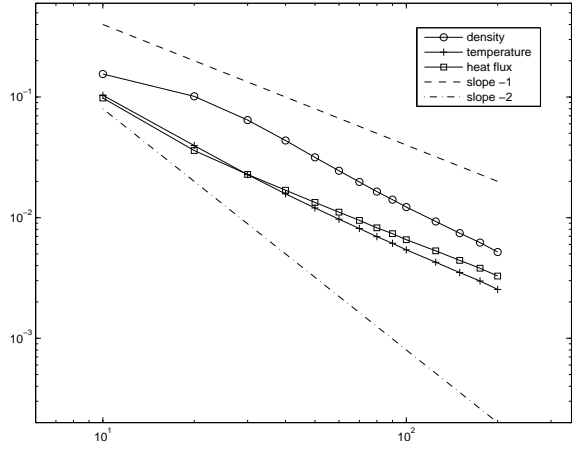}
}
\subfigure[$M=6$]{
  \includegraphics[width=.5\textwidth]{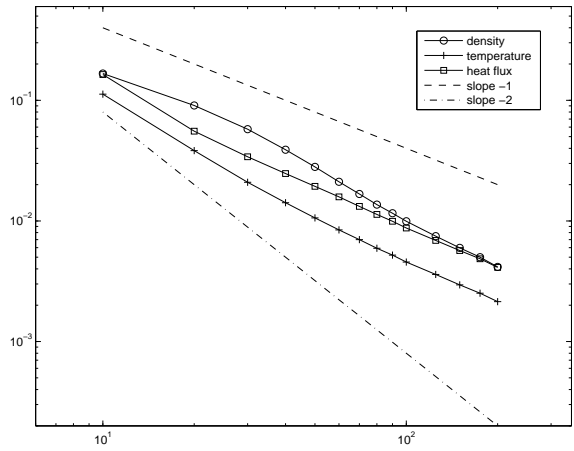}
}
\subfigure[$M=9$]{
  \includegraphics[width=.5\textwidth]{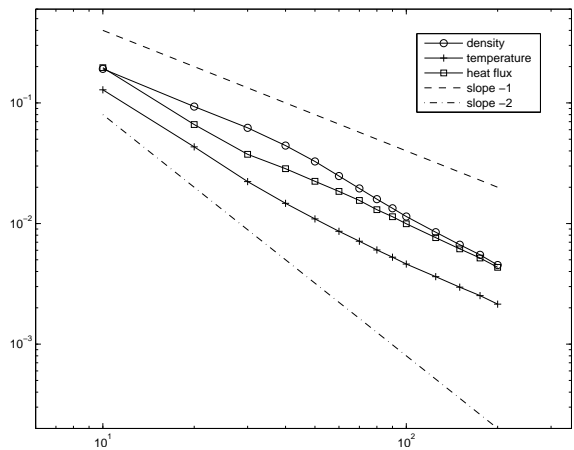}
}
\caption{Convergence rates for different moment systems}
\label{fig:periodic:convergence}
\end{figure}

Also, the first-order convergence rate is illustrated in Figure
\ref{fig:periodic:convergence}, where the ``exact'' solution is
obtained on a mesh with $800$ grids and the $L^1$ errors are shown. We
would like to remark that although the scheme is still of first order,
the magnitude of error is much smaller than the original method. In
\cite{NRxx}, where the HLL scheme without reconstruction was utilized,
the authors used $1000$ grids to obtain numerical results with similar
resolution as those in Figure \ref{fig:periodic:solution}. Note that
small grid size implies smaller time steps, which leads to a huge
computational cost.

\subsection{The shock-tube test} \label{sec:shock_tube}
In this example, we show that our method is able to achieve high
resolution when sharp layers exist in the numerical solution. Here a
Riemann shock-tube problem is considered. It has been studied by Yang
and Huang in \cite{Yang} using the discrete ordinate method. The
initial states are
\begin{equation}
(\rho, \bu, \theta) = \left\{ \begin{array}{ll}
(\rho_l, \bu_l, \theta_l), & x < 0.5, \\
(\rho_r, \bu_r, \theta_r), & x > 0.5
\end{array} \right.
\end{equation}
with
\begin{equation}
\begin{gathered}
\rho_l = 0.445, \quad \bu_l = (0.698\sqrt{2},0,0)^T,
  \quad \theta_l = 13.21,\\
\rho_r = 0.5, \quad \bu_r = (0,0,0)^T, \quad \theta_r = 1.9.
\end{gathered}
\end{equation}
The Knudsen number are selected by setting $\Kn' = 0.001$ in \eqref
{eq:Knudsen}. The computational domain is set to be $[0, 1]$, and we
solve the problem until $t = 0.1314/\sqrt{2} \approx 0.09291$ in order
to match the results in \cite{Yang}.

\begin{figure}[p]
\centering
\subfigure[$N = 100$]{
  \includegraphics[width=.45\textwidth]{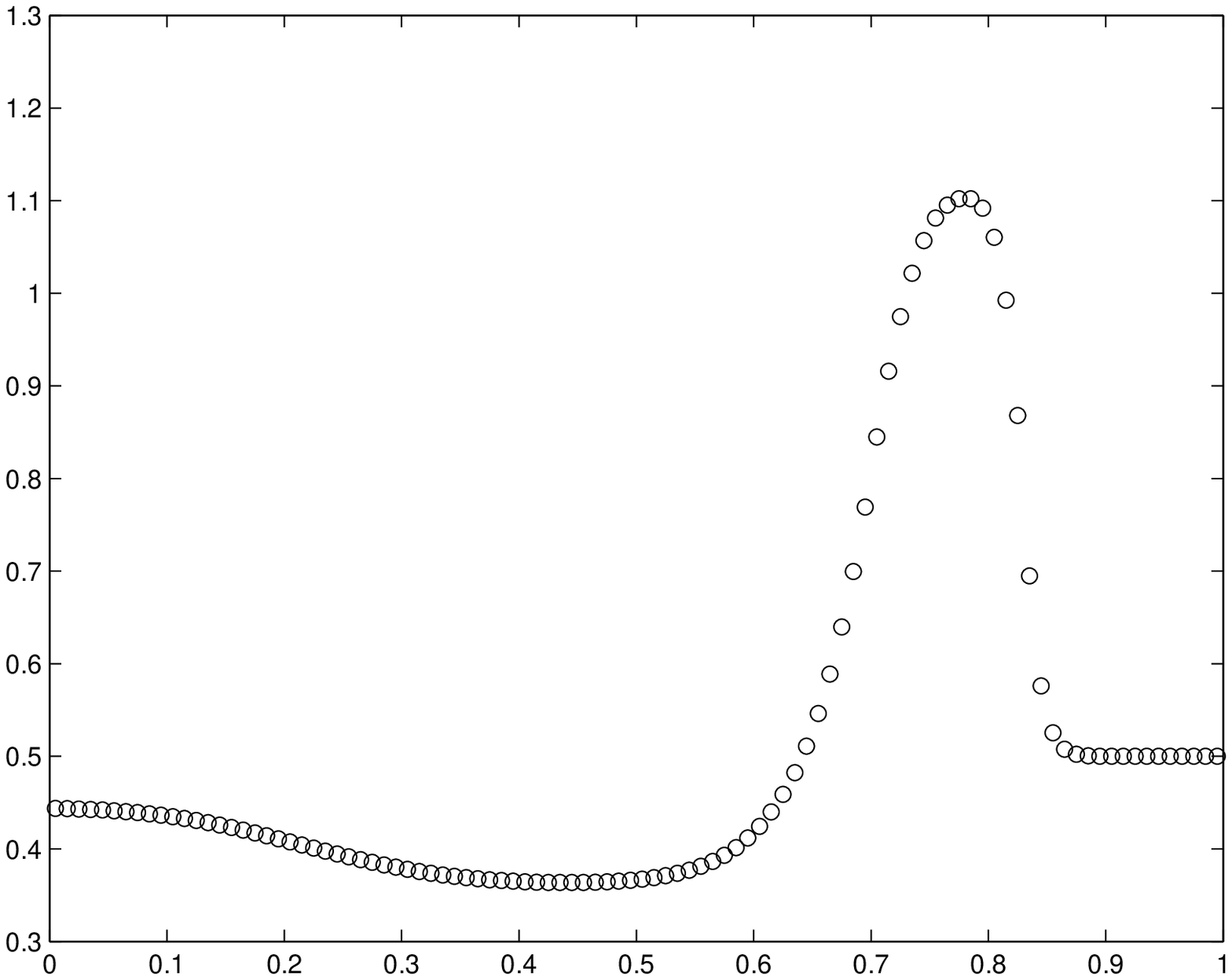}
}
\subfigure[$N = 100$ without reconstruction]{
  \includegraphics[width=.45\textwidth]{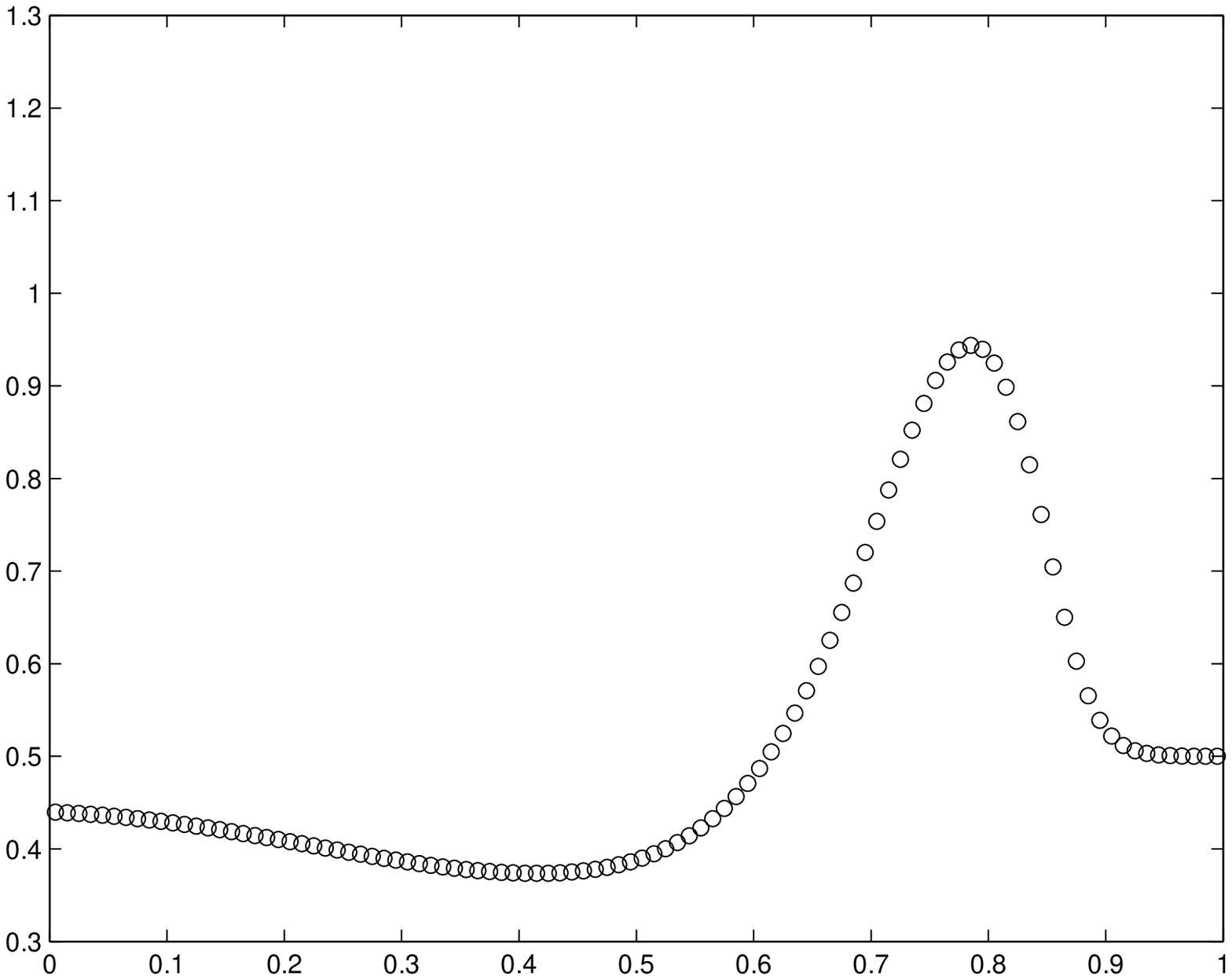}
}
\subfigure[$N = 200$]{
  \includegraphics[width=.45\textwidth]{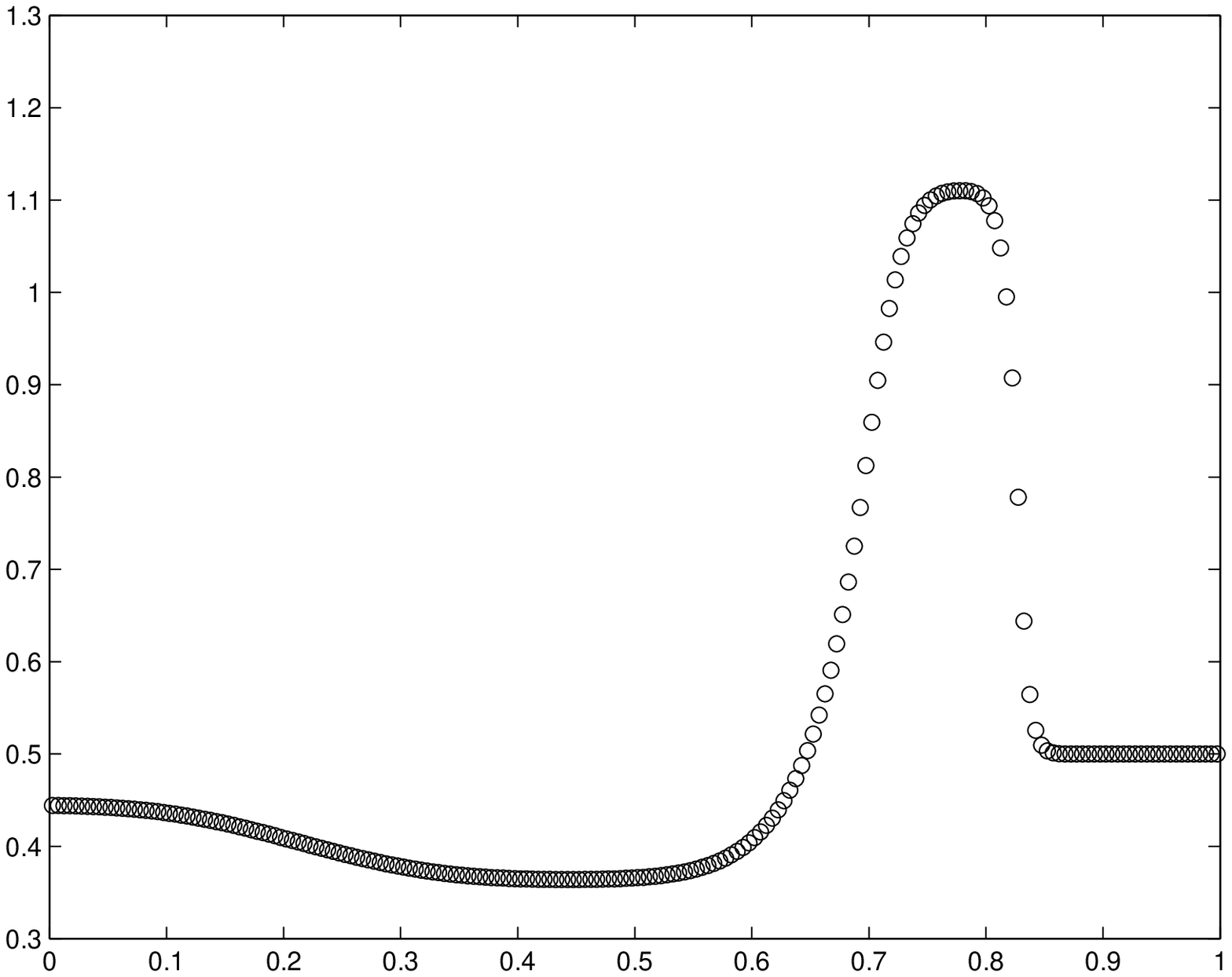}
}
\subfigure[$N = 200$, without reconstruction]{
  \includegraphics[width=.45\textwidth]{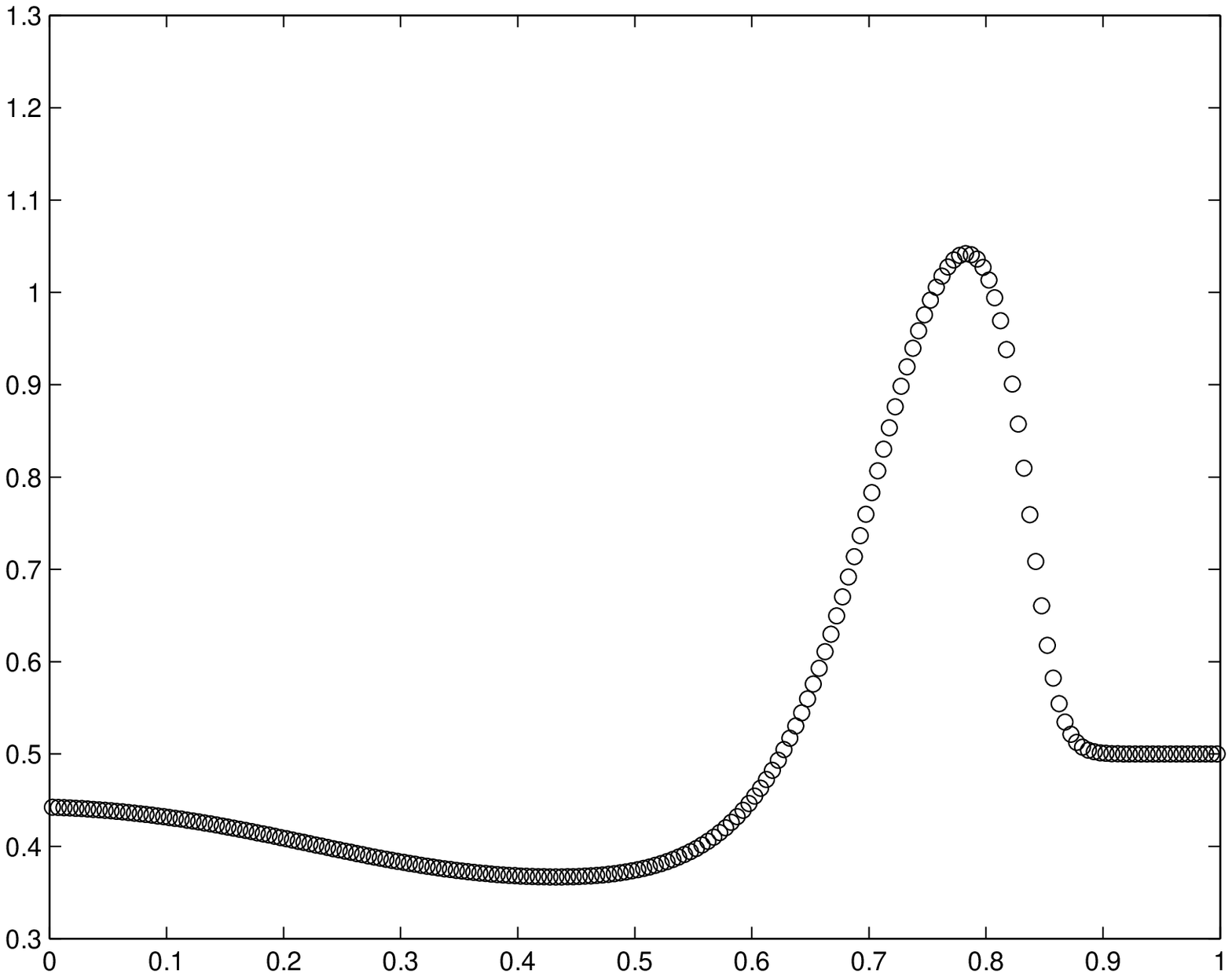}
}
\subfigure[$N = 400$]{
  \includegraphics[width=.45\textwidth]{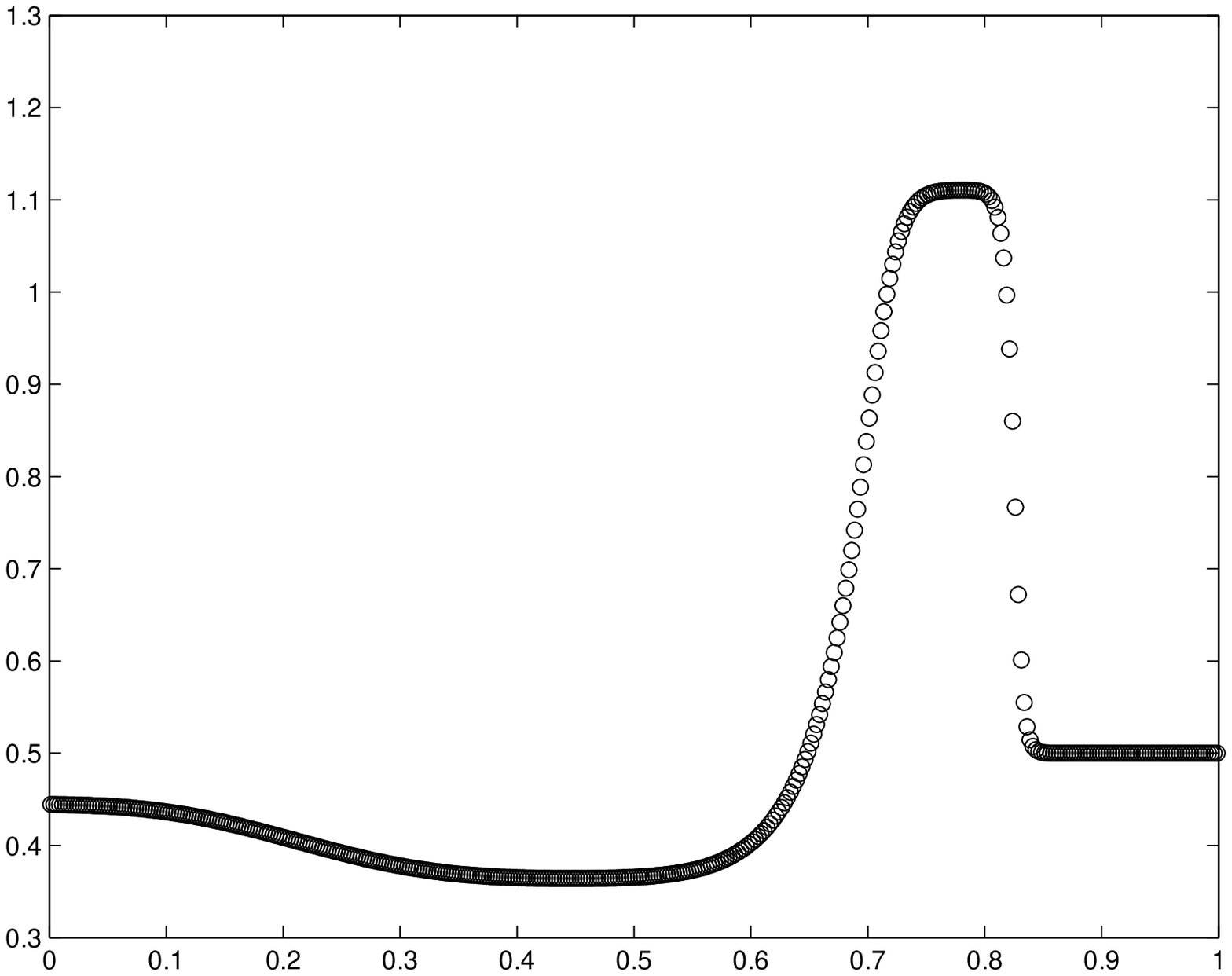}
}
\subfigure[$N = 400$, without reconstruction]{
  \includegraphics[width=.45\textwidth]{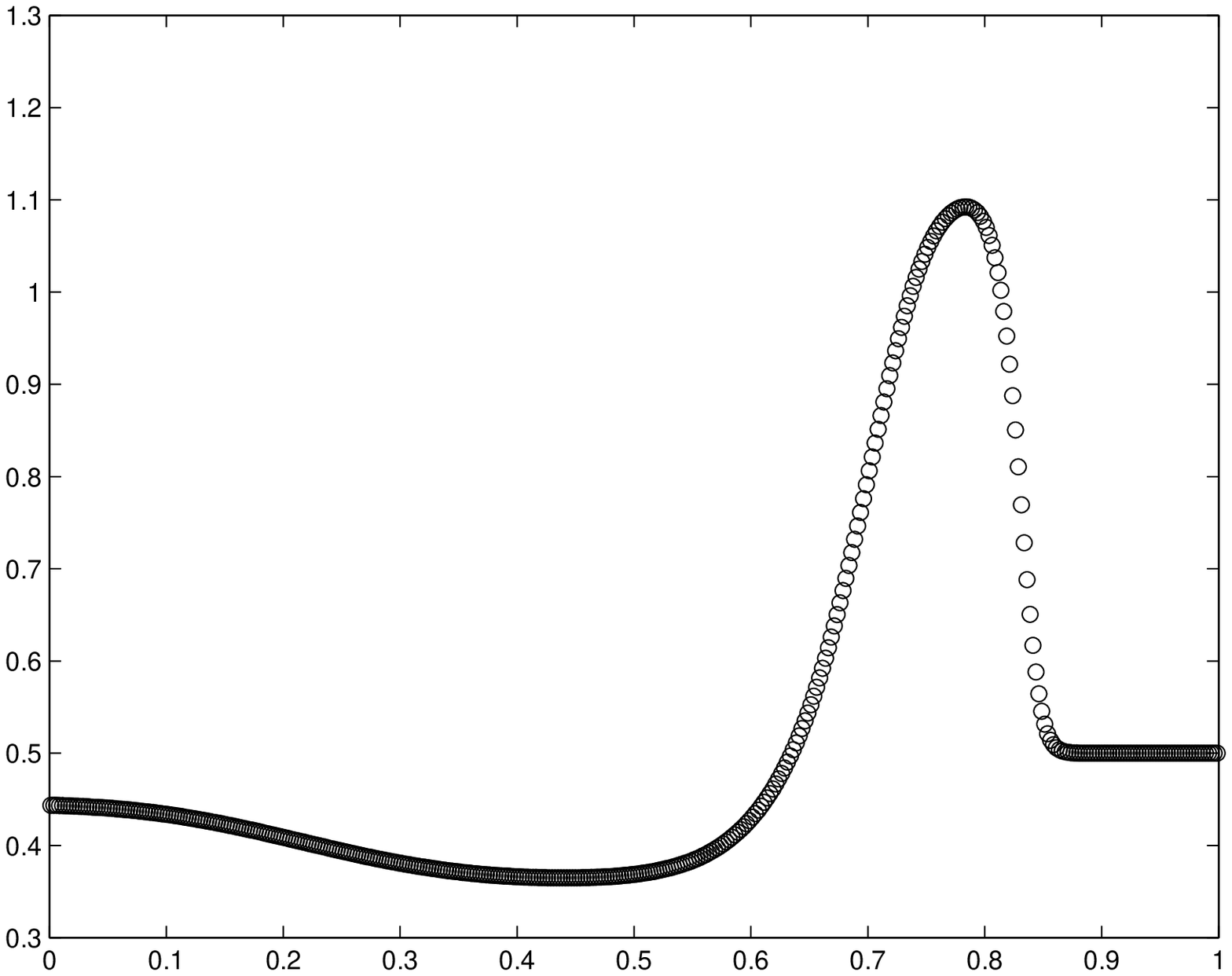}
}
\caption{Density plots of the numerical solutions of Problem
\ref{sec:shock_tube}. $N$ is the grid number.}
\label{fig:shock_tube:solution}
\end{figure}

Since $\Kn$ is small, only the case $M = 3$ is considered here. If
large $M$ is used, the results are nearly identical to the current
case. Some results are listed in the left column of Figure \ref
{fig:shock_tube:solution}, whose validity can be confirmed by comparing
them with those in \cite{Yang}. In order to see the effects of
reconstruction, we set $g_{i,\alpha} \equiv 0$ in \eqref{eq:g_i_coef}
and rerun the program. The results are in the right column of Figure
\ref{fig:shock_tube:solution}. It is obvious that the left column
provides much higher resolution near the shock wave, while the right
column is even unable to achieve the correct peak value for $N = 100$
and $N = 200$.

\section{Concluding remarks} \label{sec:conclusion} An efficient
numerical scheme with high resolution for the \NRxx method has been
presented. Since the \NRxx method gives a convective-diffusive system,
we not only perform the linear reconstruction to gain a high spatial
resolution, but also use the RKC schemes and the Strang splitting
method to enlarge the time step while maintaining the order of
accuracy in $\Delta x$. In the future work, we are extending it into
the 2D case with unstructured grids, together with the specularly
reflective boundary conditions.

\section*{Acknowledgements}
The research of the second author was supported in part by the
National Basic Research Program of China under the grant 2010CBxxxxx
and the National Science Foundation of China under the grant 10771008
and grant 10731060.


\bibliographystyle{plain}
\bibliography{../article}
\end{document}